\documentclass[a4paper,12pt]{article}
\pdfoutput=1
\usepackage{jheppub}
\usepackage{datetime}
\usepackage{float}

\begin{document}

\title{A nAttractor Mechanism for nAdS$_2$/nCFT$_1$ Holography}

\abstract{We study the nearly AdS$_2$ geometry of nearly extremal black holes in ${\cal N} = 2$ supergravity in four dimensions. 
In the strictly extreme limit the attractor mechanism for asymptotically flat black holes states that the horizon geometries of these black 
holes are independent of scalar moduli. We determine the dependence of the near extreme geometry on asymptotic moduli and express the
result in simple formulae that generalize the extremal attractor mechanism to nearly extreme black holes. This is a nAttractor mechanism.}

\author{Finn Larsen} 
  
\affiliation{Department of Physics and Leinweber Center for Theoretical Physics, \\University of Michigan, Ann Arbor, MI 48109-1120, USA.}


\maketitle


\section{Introduction}

Extremal black holes in four asymptotically flat dimensions all have near horizon geometry AdS$_2\times S^2$ \cite{Kunduri:2007vf}. It is therefore 
important to understand the AdS$_{d+1}$/CFT$_d$ correspondence in the special case $d=1$ where the holographic correspondence is
between AdS$_2$ quantum gravity and CFT$_1$ conformal quantum mechanics. Unfortunately, this case has proven very different from examples of holography with other values of $d$ and it remains poorly understood. 

The special nature of AdS$_2$ manifests itself in classical gravity where asymptotically AdS$_2$ boundary conditions preclude finite energy 
excitations in the interior of spacetime, rendering strict AdS$_2$ gravity nondynamical \cite{Maldacena:1998uz,Almheiri:2014cka,Engelsoy:2016xyb}. It is also evident in considerations of the decoupling limits of string theory that motivate holography with $d= 2, 3, 4, 6$ \cite{Maldacena:1997re,Itzhaki:1998dd} where generalizations to $d=1$ fail \cite{Maldacena:1998uz}. 
Such results suggest that dynamics 
localized in the AdS$_2$ near horizon geometry necessarily couple to modes further away from the horizon. 

In the last few years a precise version of AdS$_{2}$/CFT$_1$ duality was developed that confronts these obstacles in a straightforward manner: 
it incorporates modes that have support beyond the AdS$_2$ horizon geometry, but only ``near" this region. Such modes also introduce a small 
amount of energy that violates the extremal limit, rendering the corresponding black holes ``nearly" extremal. Moreover, these small deformations are inconsistent with conformal symmetry, but the theory remains ``nearly" conformal. The configuration space of such deformations yields a construction that is interesting because the dynamical theory of the excitations near the extremal limit is thought to be holographic: the 2D bulk theory with gravity is dual to a 1D boundary theory without gravity. This duality is referred to as the nAdS$_{2}$/nCFT$_1$ correspondence \cite{Maldacena:2016hyu,Maldacena:2016upp,Jensen:2016pah,Kitaev:2017awl}

Effective quantum field theory is an excellent tool for analyzing this situation. In this language the AdS$_2$ geometry is interpreted as a highly degenerate ground state that is protected by a vast symmetry that renders it non-dynamical, possibly topological. This CFT$_1$ features irrelevant operators that, when engaged with finite coefficient, deform the theory so that it becomes dynamical. The resulting theory has access only to a small part of the entire phase space and it breaks scale invariance. The power of effective quantum field theory is that the low energy description is determined largely by the broken symmetry. It has stimulated much research that theories with this symmetry breaking pattern are realized by numerous quantum mechanical theories that are simple enough that they can be analyzed in significant detail, specifically the SYK model and its avatars \cite{Maldacena:2016hyu,Gross:2016kjj,Witten:2016iux,Klebanov:2016xxf,Sarosi:2017ykf}.

The purpose of this paper is to develop the gravitational side of the proposed duality in the context of asymptotically flat (nearly) BPS 
black holes in ${\cal N}=2$ supergravity. These black holes realize a cornucopia of explicit nAdS$_2$ geometries that have been analyzed in detail from many different points of view. One of their important aspects is the attractor behavior they exhibit in the strict extremal limit: the values of scalar fields at the horizon depend only on the black hole charges, they are independent of the boundary conditions on scalar fields in the asymptotically flat space far from the black hole \cite{Ferrara:1996dd,Strominger:1996kf}. We will generalize this result and demonstrate that the {\it near} horizon field configuration of a {\it nearly} extreme black hole follows from the {\it extremal} attractor mechanism by a simple rule:  it is determined by the {\it derivatives} of the corresponding attractor values for extremal black holes with respect to the {\it conserved charges}. The direction of the gradient in the space of charges is specified by the asymptotic values of moduli. This algorithm offers a powerful and convenient way to obtain near horizon values of fields without constructing the entire black hole solution. We refer to these results as a nAttractor mechanism. 

Quantum gravity in nAdS$_2$ is not scale invariant because the dilatation symmetry of AdS$_2$ fails beyond the horizon region
of the extreme black hole. The scale (or scales) thus acquired by the theory is of fundamental interest. The nAttractor mechanism offers a simple method to compute such symmetry breaking scales precisely.

The fundamental reference scale for AdS$_2$ geometry is the curvature scale $\ell_2$ (which is identical to the radius of the $S^2$). However, as in any AdS$_{d+1}$/CFT$_d$ correspondence this length is not an intrinsic scale, it is just a unit. For example, a minimally coupled scalar field of mass $m$ propagating in the AdS$_{d+1}$ background is characterized by the dimensionless conformal weight 
\begin{equation}
\label{eqn:hform}
h = {d\over 2} + \sqrt{{d^2\over 4}+ m^2\ell^2_{d+1}}~.
\end{equation}
The scale $\ell_{d+1}$ enters all aspects of the theory only as a means of forming dimensionless quantities such as $m\ell_{d+1}$. 

In the context of extreme black holes with AdS$_2\times S^2$ horizon geometry, a genuine physical scale characterizing AdS$_2$ quantum gravity is introduced by $G_4$, Newton's gravitational constant in 4D. A good measure is the entropy of the 4D extremal black hole
\begin{equation}
S_0 = {4\pi \ell^2_2\over 4 G_4}  = {1\over 4 G_2} ~, 
\label{eqn:s0def}
\end{equation}
where $G_2$ is the gravitational constant in 2D. We presume that it is meaningful to describe the black hole by classical gravity in some approximation so the dimensionless constant $G_2$ must be very small or, equivalently, the ground state entropy \eqref{eqn:s0def} must be huge. This reflects the fact that the 4D Planck scale is much smaller than the AdS$_2$ scale $\ell_2$ and generally too small to be useful for the semi-classical description. 

The {\it near} horizon region of {\it nearly} extreme black holes is qualitatively different from the horizon region of the extremal black hole because it involves a new classical length scale. In the gravitational representation the scale invariant AdS$_2$ throat ``bends over" due to a deformation that is small and therefore characterized by a length scale that is much larger than $\ell_2$. We will find that the new scale is characterized efficiently by the radial derivative of fields in the AdS$_2$ geometry. In the language of the dual quantum theory the near horizon geometry is characterized by one or more irrelevant operators with small but finite coefficients, in the manner formalized by conformal perturbation theory. In the setting we study all these operators will have $m^2\ell^2_2=2$ which, according to \eqref{eqn:hform} with $d=1$, corresponds to conformal dimension $h=2$. Importantly, the ``small" coefficients introduced this way are dimensionfull and their smallness means they are characterized by length scales that are 
much larger than $\ell_2$, as they must be when the operators are irrelevant. 

A robust way to introduce a length scale that characterizes the departure from the extremal limit is through the specific heat 
$C_{P,Q}$ at low temperature \cite{Preskill:1991tb}: 
\begin{equation}
L = {2\over\pi} {C_{P,Q}\over T}=  {2\over\pi} \left({\partial S\over\partial T}\right)_{P,Q}~.
\end{equation}
The specific heat is proportional to the absolute temperature as $T\to 0$ so the length scale $L$ is finite in the limit. The indices $P,Q$ indicate that the temperature is lowered with magnetic and electric charges kept fixed. The awkward numerical factor ${2\over\pi}$ is chosen so that $L$ coincides with the long string scale \eqref{eqn:longstr} in the dilute gas limit of the STU-model. It is equivalent to the mass gap $M_{\rm gap}= L^{-1}$. 

Qualitative reasoning based on the scales already introduced suggests a ``typical" value for the symmetry breaking scale
\begin{equation}
\label{eqn:genericL}
L_{\rm typ} = {8\pi\ell^3_2\over G_4} =  8 \ell_2 S_0 ~.
\end{equation}
This length is much larger than the AdS$_2$ scale $\ell_2$ because the the ground state entropy $S_0$ is huge. It has been conjectured that the effective quantum field theory of AdS$_2$ quantum gravity is universal in the infrared limit, in the sense that it is characterized entirely by a single length scale, the symmetry breaking scale $L$ \cite{Almheiri:2016fws,Maldacena:2016upp}, which may be estimated by $L_{\rm typ}$.  The central point of this article is that many theories of interest have scalar moduli and generally the {\it symmetry breaking scale depends on these continuously tunable parameters}. Equivalently, we may refer to multiple scales in the theory.

Specifically, ${\cal N}=2$ supergravity with $n_V$ vector multiplets generically features a total of $2n_V+1$ scales. We could identify the ``typical" scale \eqref{eqn:genericL} as the scale associated with the scalar field that represents the size of the spatial sphere in the 2D theory. Then the $2n_V$ additional scales that appear in the generic situation are interpreted from the UV perspective as standard moduli, ie asymptotic data for scalar fields $z^i$ imposed in the flat space far from the black hole. Generally these scalar fields all source gravity in the near horizon region so there are many scales and there is no operational sense in which there is only one scale.

The much studied Reissner-Nordstr\"{o}m black hole arises as a finely tuned special case where all scales vanish except for one that is precisely the typical scale
\eqref{eqn:genericL}. Another important special case is when the AdS$_2$ theory can be interpreted as dimensional reduction of an AdS$_3$ theory. These ``dilute gas" black holes \cite{Maldacena:1996ix} correspond to all scales taking the same value so effectively there is again just one scale. However, it is much larger than the ``typical" one \eqref{eqn:genericL}. 

In the interest of clarity, this paper will consider only the simplest setting: asymptotically flat, spherically symmetric black holes in 4D with a smooth BPS limit, as solutions to standard two-derivative  ${\cal N}=2$ SUGRA supergravity with matter in the form of vector multiplets \cite{Ferrara:1995ih}. The attractor mechanism for black holes has been fruitfully generalized to many other situations, including features such as multicenter solutions \cite{Denef:2007vg}, nonsupersymmetric black holes \cite{Goldstein:2005hq}, extended SUGRA \cite{Ferrara:2006em}, rotation \cite{Astefanesei:2006dd,Castro:2018}, higher derivative corrections \cite{LopesCardoso:1998tkj,Ooguri:2004zv,Castro:2007sd}, asymptotically AdS spacetime \cite{Morales:2006gm,Astefanesei:2007vh,Cacciatori:2009iz}, five dimensions \cite{Larsen:2006xm}, and extended black objects \cite{Kraus:2005gh}. The discussion in this paper can presumably be generalized to most or all of these settings and explicit study of such examples would cast more light on the universality classes of 2D quantum gravity. 

This paper is organized as follows. In section 2 we briefly review ${\cal N}=2$ SUGRA in 4D. An appendix on special geometry and another on the equations of motion offer some more details. Section 3 is the central part of the paper: we show that the form of departures away from the extremal limit of black holes is characterized by the extremal limit itself but the coefficients of such deformations are free parameters that can be interpreted as asymptotic data. The result can be cast as a simple formula for all symmetry breaking scales of the near horizon theory. In section 4 we make the results explicit in the special case of STU black holes with charges limited to the ``four charge" configuration.

\section{${\cal N}=2$ Supergravity and its 2D Reduction}
\label{sec:eom}
In this section we introduce ${\cal N}=2$ SUGRA in 4D and its spherical reduction to 2D in a charged black hole background. 
This section also serves as a concise introduction to notation and terminology. Some additional details of special geometry are reviewed 
in Appendix \ref{app:spec}.

\subsection{Spherical Reduction of 4D {\cal N}=2 Supergravity}

The bosonic matter of 4D ${\cal N}=2$ supergravity coupled to $n_V$ vector multiplets comprises the metric $g_{ab}$, $n_V+1$ vector fields $A_a^{I}$ (with $I=0,1,..,n_V$), and $n_V$ complex scalar fields $z^i$ (with $i=1,..,n_V$). They are described by the 4D Lagrangian
\begin{eqnarray}
16\pi G_4 {\cal L}_4 & = & {\cal R}^{(4)} - 2 g_{i{\bar\j}} \partial_a z^i \partial^a {\bar z}^{\bar\j} + {1\over 2} \nu_{IJ}F^I_{ab} F^{Jab} +  
{1\over 2}\mu_{IJ}F^I_{ab} {}^*F^{Jab} ~.
\label{eqn:4dlag}
\end{eqnarray}
The K\"{a}hler metric $g_{i{\bar\j}}$ and the gauge kinetic functions $\mu_{IJ}$, $ \nu_{IJ}$ depend on the scalar fields through the rules of special geometry reviewed in Appendix \ref{app:spec}. We do not include hypermultiplets in the Lagrangian because the hypermoduli decouple from other fields. Thus the attractor mechanism applies only to K\"{a}hler moduli (complex structure moduli) in compactifications of type IIA (type IIB) string theory on a Calabi-Yau threefold. 

We want to analyze solutions to the 4D theory \eqref{eqn:4dlag} from a 2D point of view. An appropriate {\it ansatz} for the 4D metric is
$$
ds^2_4 = g_{\mu\nu} dx^\mu dx^\nu + R^2 d\Omega^2_2~,
$$
where now $g_{\mu\nu}$ is a 2D metric and the radial function $R^2$ depends on the 2D base only. Dimensional reduction of the 4D Lagrangian \eqref{eqn:4dlag} to 2D is straightforward, except for the gauge fields which require some care, as follows. 

The gauge field strengths satisfy their Bianchi identities and their equations of motion 
\begin{eqnarray}
\nabla^a {}^*F^I_{ab} & = & 0 ~, \cr
\nabla^a  \left( \nu_{IJ}F^J_{ab}  +  \mu_{IJ} {}^*F^J_{ab}   \right)  &=&  0~. 
\end{eqnarray}
These equations can be integrated to define conserved magnetic and electric charges
\begin{eqnarray}
P^I  &=&  {1\over 4\pi} \int_{S^2} F^I ~,\cr
Q_I  &=& {1\over 4\pi} \int_{S^2}  \left( \nu_{IJ}F^J  +  \mu_{IJ} {}^*F^J  \right) ~.
\label{eqn:chargedef}
\end{eqnarray}
Since the matrices $\mu_{IJ}$, $ \nu_{IJ}$ depend on the scalar fields, it is only precisely these linear combinations of field strengths that yield 
conserved charges. 

The flux integrals \eqref{eqn:chargedef} define charges that are normalized as the coefficient of the $1/r$ term in the 
electric or magnetic potential and have dimension of length. It is often better to employ dimensionless charges 
$(p^I, q_I)= (P^I, Q_I)/\sqrt{2G_4}$ that are quantized and can be identified with the corresponding number of 
constituent ``branes" with the given type of charge. 

For spherically symmetric black holes the definitions  \eqref{eqn:chargedef} easily determine the two-forms $F^I$ in terms of the charges $(P^I, Q_I)$, which 
can then be substituted into the action \eqref{eqn:4dlag}. For the magnetic components (with indices of $F^I$ on the sphere) this is the correct 
procedure but for the electric components (with indices of $F^I$ on the AdS$_2$ base) it is not. To keep the conserved electric charge fixed in the spherical reduction we
must Legendre transform ${\cal L}_2\to {\cal L}_2 - Q_I F^I$ (the Lagrangian density is treated as a two form in this substitution). 
The net effect of this procedure is that the sign of the 4D gauge kinetic terms flips
and so the electric charges contribute a positive term to the effective 2D potential, just like the magnetic charges.\footnote{This maneuver is common in classical mechanics. It is needed to address a charge that is conserved because it is conjugate to a cyclic variable. According to standard terminology in this context it would be
appropriate to refer to ${\cal L}_2$ given in \eqref{eqn:2dlag} as (the negative of) the {\it Routhian} \cite{Landau:1986ao}. We will forego this nomenclature because, for our purposes, ${\cal L}_2$ can be employed as a conventional Lagrangian.}

Having taken gauge fields properly into account, the effective 2D action for 4D ${\cal N}=2$ SUGRA becomes
\begin{eqnarray}
4G_4 {\cal L}_2 &=& R^2 {\cal R}^{(2)} + 2 + 2(\nabla R)^2 - 2 R^2 g_{i{\bar j}} \nabla_\mu z^i \nabla^\mu {\bar z}^{\j} - {2V_{\rm eff} \over R^2} ~,
\label{eqn:2dlag}
\end{eqnarray}
where the 2D effective potential is 
\begin{equation}
\label{eqn:Vpot}
V_{\rm eff} =  G_4  
\left( p^I ~q_I \right) 
\left(  
\begin{matrix} \nu_{IJ} + \mu_{IK} (\nu^{-1})^{KL}\mu_{LJ} && -\mu_{IK} (\nu^{-1})^{KJ}    \\ - (\nu^{-1})^{IK}\mu_{KJ}  
&& (\nu^{-1})^{IJ}    
\end{matrix}
\right)
\left(  
\begin{matrix} 
  p^J \\ q_J 
  \end{matrix}
\right)
~.
\end{equation}

The dynamics encoded in the dependence of the gauge kinetic functions $\mu_{IJ}, \nu_{IJ}$ on the scalar fields can be quite elaborate. 
For extended SUGRA (${\cal N}>2$ SUSY) the scalar manifolds are always cosets $G/H$ of (semi-)simple groups and this connection 
can give manageable parametrizations of the effective potential \eqref{eqn:Vpot}. Here the focus is on general ${\cal N}=2$ SUGRA 
and then it is convenient to parametrize the scalar fields as a symplectic pair $(X^I, F_I)$ and introduce the spacetime central charge 
\begin{equation}
{\cal Z} = {1\over\sqrt{G_4}} e^{{\cal K}/2}\left( X^I q_I - F_I p^I\right)~.
\label{eqn:Zdef}
\end{equation}
The symplectic parametrization of the scalar fields is projective so only the ratios $z^i=X^i/X^0$ (with $i=1,\ldots,n_V$) are physical. 
The K\"{a}hler potential ${\cal K}$ \eqref{eqn:kahlerdef} compensates for the redundancy of the projection and renders the central 
charge ${\cal Z}$ physical. 
Indeed, the central charge appears in the ${\cal N}=2$ SUSY algebra such that the BPS black hole mass becomes $M=|{\cal Z}|_\infty$ 
with the index ``$\infty$" indicating evaluation of the moduli and the K\"{a}hler potential in the asymptotically flat space. The central charge generally depends on spacetime location and, for any ${\cal N}=2$ SUGRA, it is possible to recast the effective 2D potential \eqref{eqn:Vpot} in terms of the spacetime central charge as
\begin{equation}
V_{\rm eff} =  G_4 \left( |{\cal Z}|^2 + 4g^{i{\bar\j}}\partial_i |{\cal Z}| \bar{\partial}_{\bar\j} |{\cal Z}| \right)~,
\end{equation}
where $\partial_i$ are partial derivatives with respect to the scalar fields $z^i$. This form simplifies the equations of motion. 

\section{Near Horizon Geometry of Nearly BPS Black Holes}
\label{sec:bhsols}
In this section we we first consider general  features of spherically symmetric black holes ${\cal N}=2$ SUGRA in 4D. 
We show that the near horizon geometry of near extreme black holes can
be extracted from the near horizon region of extremal solutions. We then exploit results from the literature on BPS black holes 
to find simple formulae for the symmetry breaking scales of the near BPS black holes. 

\subsection{Nonextremal Black Holes in ${\cal N}=2$ SUGRA}
\label{sec:nextbhs}
All spherically symmetric black holes in 4D can without loss of generality be written in the form 
\begin{equation}
\label{eqn:gennnext} 
ds^2_4 =  - e^{2\Phi} dt^2 +   e^{-2\Phi} dr^2 + R^2 d\Omega^2_2~,
\end{equation}
where $\Phi, R$ are functions of the radial coordinate $r$. It is straightforward (but not terribly illuminating) to find the equations of 
motion for $\Phi, R$ that follow from 
the 2D Lagrangian \eqref{eqn:2dlag}. An important general result that follows from these manipulations is that $\Phi$ and $R$ must be related 
so that we can write the 4D metric \eqref{eqn:gennnext} in terms of a single radial function $R$ as 
\begin{equation}
\label{eqn:gennonext} 
ds^2_4 =  -{r^2-r^2_0\over R^2} dt^2 + {R^2\over  r^2-r^2_0} dr^2 + R^2 d\Omega^2_2~,
\end{equation}
where $r_0$ is a constant of integration that parametrizes the departure from extremality of the black hole. This
result applies to all black holes, not necessarily near extremality\footnote{Equivalent results were 
reported in \cite{Gibbons:1982ih,Gibbons:1996af,Ferrara:1997tw}
($c^{\rm there} = 2r_0^{\rm here}$ and $G^{\rm there}_4=1$) and elsewhere.}. We present an elementary derivation in Appendix \ref{app:eom}.

We can establish some features of the geometry \eqref{eqn:gennonext} for a radial function $R^2(r)$ that is arbitrary, except that we assume regularity at $r=r_0$. For example, we can compute the black hole entropy from the area law applied to the geometry \eqref{eqn:gennonext} 
at the event horizon $r=r_0$
\begin{equation}
\label{eqn:bhentropy} 
S = {\pi R^2(r_0) \over G_4}~.
\end{equation}
We can also compute the black hole temperature by imposing regularity of the Euclidean continuation of the Lorentzian 
geometry at $r=r_0$. It becomes
\begin{equation}
T = {r_0\over 2\pi R^2(r_0)}~.
\label{eqn:Tmrel}
\end{equation}
The product of these two equations give the relation 
\begin{equation}
G_4 TS = {r_0\over 2}~,
\label{eqn:GTSm}
\end{equation}
for any black hole of the form \eqref{eqn:gennonext}. In particular, the previous three equations do not rely on any kind of near extreme limit. 

\subsection{The Approach to the Extremal Limit}
To define the extremal limit precisely, recall that thermodynamic variables such as the entropy $S$ and the temperature $T$ depend on 
the black hole state parameters, like the mass $M$, charges $(P^I,Q_I)$, and also on the vacuum of the theory specified by the scalars at infinity $z^i_\infty$. The extremal limit lowers the temperature $T\to 0$ by lowering the mass $M\to M_{\rm ext}$ {\it with all other parameters fixed}. 

According to this prescription the entropy changes as the extremal limit is approached, because it depends on the mass $M$. However, the radial function $R^2(r)$ changes {\it both} because it depends on the mass parameter $M$ and {\it also} because the horizon moves, as measured by its coordinate position. All these changes are related as
\begin{equation}
\Delta S = {\partial S\over \partial M}  \Delta M = {\pi\over G_4} \left( {\partial R^2\over \partial M} \Delta M + {\partial R^2\over \partial r} \Delta r\right) ~,
\label{eqn:delS}
\end{equation}
where we used \eqref{eqn:bhentropy} for the entropy. 

The black entropy approaches the extremal limit smoothly $S\to S_{\rm ext}$ with no discontinuity. It follows from \eqref{eqn:GTSm} that the extremal limit $T\to 0$ is equivalent to $r_0\to 0$ with $T$ and $r_0$ {\it proportional}. The coordinate position of the horizon changes by $\Delta r=r_0$ in the final approach to extremality so we find that the last term in \eqref{eqn:delS} is {\it linear} in the temperature $T$. 

The first law of thermodynamics, expressed by the first equation in \eqref{eqn:delS} with $\partial_M S = T^{-1}$, shows that a change in the entropy that is linear with temperature $\Delta S \propto T$ corresponds to a change in mass that vanishes as the temperature {\it squared}
\begin{equation}
\Delta M=M-M_{\rm ext}\sim T^2~. 
\label{eqn:DelMT2}
\end{equation}
Therefore, the penultimate term in \eqref{eqn:delS} is negligible because it depends {\it quadratically} on the temperature. 

The preceding estimates assume that the radial function $R(r)$ has a smooth extremal limit, both as the black hole 
mass $M\to M_{\rm ext}$ at fixed coordinate $r$ and also as the radial parameter $r=r_0\to 0$ at fixed mass. The extremal limit of black holes is generally quite subtle and smoothness cannot be taken for granted. For example, the topology of the extremal black hole geometry in Euclidean signature is a cylinder (times the horizon $S^2$) while the corresponding non-extreme black hole geometry is topologically a disc (times the
horizon $S^2$) for any non-vanishing excitation above extremality. Thus there is a discontinuity at extremality. Some physical aspects of the extreme limit are discussed in \cite{Preskill:1991tb,Carroll:2009maa}. 

Fortunately, the smoothness that is needed here is fairly mild: the function $R^2$ and its derivatives $\partial_r R^2$, $\partial_M R^2$ must be finite and continuous in the extreme limit. The metric was presented in the general form \eqref{eqn:gennonext} in order to satisfy these conditions. For the familiar Reissner-Nordstr\"{o}m black holes $R=r+M$ is manifestly regular in both variables and it is straightforward that the conditions are similarly satisfied in more elaborate explicit solutions (such as the STU-black holes \cite{Cvetic:1995uj}). As a general argument, we note that the metric \eqref{eqn:gennonext} depends only on the variable 
$r_0^2\sim T^2 \sim M-M_{\rm ext}$. Therefore, we expect that the equations of motion \eqref{eqn:simpleom} can be solved as a perturbation series in $M-M_{\rm ext}$ around the extremal solution. This expansion will always give a finite value for $\partial_M R^2$. 

It follows from the estimates after \eqref{eqn:delS} that we can express the entropy due to the departure from extremality as 
\begin{equation}
\Delta S = S - S_{\rm ext}  = {\pi\over G_4} {\partial R^2\over \partial r} \Delta r= {\pi r_0\over G_4} {\partial R^2(0)\over\partial r}~,
\end{equation}
to the leading order. It corresponds to the symmetry breaking scale
\begin{equation}
L =  {2\over\pi} {\Delta S \over T} = {2\pi\over G_4}  {dR^4(0)\over dr}~,
\label{eqn:Lformula}
\end{equation}
since the temperature is given by \eqref{eqn:Tmrel}. The expression \eqref{eqn:Lformula} is remarkable because 
the left hand side is a property of {\it near}-extremal black holes that we seek to compute while the right hand 
side {\it depends only on the BPS solution}. Thus the ``near" in the near extremal limit is immaterial in this context and the ``near" in 
near horizon is incorporated in the simplest possible fashion, as a derivative in the coordinate normal to the horizon. 

We can use the analogous reasoning on the complex scalar fields $z^i$ in vector multiplets of ${\cal N}=2$ supergravity. 
Like the radial function $R$, these fields are functions of the coordinate
$r$ and they also depend on the black hole state parameters. As we approach the extremal limit the scalar fields therefore change 
due to these dependencies as
\begin{equation}
\Delta z^i ={\partial z^i\over\partial M}\Delta M + {\partial z^i\over\partial r}\Delta r~.
 \end{equation}
The scalar fields depend smoothly on the radial coordinate as well as the black hole mass. Therefore, 
the estimates around \eqref{eqn:DelMT2} establish that the first term on the right hand is negligible, the
second term is the dominant one. We conclude that the near horizon behavior of the scalar fields in a near 
extreme black hole solution can be computed from their radial dependence in the corresponding
extreme black hole. 

\subsection{Geometry of the BPS Black Hole: the Horizon Attractor}
For BPS black holes the dependence of the radial function $R^2$ and the scalars on the radial coordinate $r$ is given by the standard attractor flow. The starting point is
the supersymmetry conditions on BPS black holes which yield the first order flow equations\footnote{A common and slightly more economical form of these equations use the field $U$ and the coordinate $\tau$ \cite{Ferrara:1997tw,Bates:2003vx}, related to the variables here through $R=re^{-U}$ and $\tau=-1/r$. The field $U$ is not optimal in this work because it is not regular at the horizon.}
\begin{eqnarray}
G_4 | {\cal Z}| &=& - r^2 \partial_r \left( {R\over r}\right) = R - r\partial_r R~, \cr
G_4 \partial_i | {\cal Z}| &=&  2r R g_{i{\bar\j}} \partial_r z^{\bar\j}~.
\label{eqn:attractorflow}
\end{eqnarray}
These equations guarantee that the equations of motion \eqref{eqn:simpleom} of ${\cal N}=2$ SUGRA are all satisfied. 

The simplest solutions to the flow equations \eqref{eqn:attractorflow} describe the AdS$_2\times S^2$ attractor geometry with 
constant moduli $z^i$ . We position the attractor at $r=0$ and then the second flow equation demands that these constant scalars take values such that the spacetime central charge ${\cal Z}$ introduced in \eqref{eqn:Zdef} is extremized as moduli vary
\begin{equation}
\partial_i | {\cal Z}|=0~.
\label{eqn:DiZ}
\end{equation}
The first flow equation then shows that the value of $|{\cal Z}|$ at its extremum is identical to the constant value of $R$ that characterizes the dimension 
of the attractor geometry 
\begin{equation}
R =  G_4 | {\cal Z}|_{\rm hor}~, 
\label{eqn:RZ}
\end{equation}
except for the trivial factor $G_4$ that addresses dimensionality. The extremization principle \eqref{eqn:DiZ} with 
subsequent computation of the radial function through \eqref{eqn:RZ} is a powerful and convenient implementation of the attractor 
mechanism. The constant radius $R$ is equivalent to the area of the $S^2$ so it yields, in particular, the black hole entropy 
through \eqref{eqn:bhentropy} (with $r_0=0$ for the extremal case). 

The attractor mechanism for extremal black holes is often expressed in 
other ways. For example, ``the" attractor equations 
\begin{eqnarray}
P^I & = & G_4 {\rm Re} \left[ {\cal C} X^I\right]_{\rm hor} ~, \cr
Q_I  & = &  G_4 {\rm Re} \left[ {\cal C} F_I\right]_{\rm hor}  ~,
\label{eqn:attractpq}
\end{eqnarray}
where ${\cal C}$ is an arbitrary constant, are equivalent to the extremization principle \eqref{eqn:DiZ} but they are purely algebraic. These equations have
been much studied and general aspects of their solutions are well understood. 

In general, the function $R$ is a function of the radial coordinate, with charges $(P^I, Q_I)$ and moduli $z_\infty^i$ appearing as parameters. At an extremal horizon the
radial coordinate is fixed at $r=0$ and the attractor behavior amounts to independence of $z_\infty^i$; so $R$ depends only on black hole charges. 
The charges have the same dimension as length so $R$ will be a homogenous function of degree one. It is conventional to present it as 
\begin{equation}
R^4(0) = I_4(P^I, Q_I)~,
\label{eqn:R4att}
\end{equation}
where $I_4(P^I, Q_I)$ is a homogenous function of degree four. This notation is motivated by the result that $I_4$ is a quartic polynomial in the charges for most widely studied explicit examples. It is the linchpin of any solution to the attractor equations because, given $I_4(P^I, Q_I)$, the values of the scalar fields in the attractor geometry can be expressed in terms of the same function \cite{Bates:2003vx}. Presenting the scalars as a symplectic pair $(X^I, F_I)$, the attractor values are
\begin{equation}
\label{eqn:PQtoXFmap}
\left(  
\begin{matrix} 
  X^I_{\rm hor} \\ F^{\rm hor}_I 
  \end{matrix}
\right) = 
\left(  
\begin{matrix} 
  P^I \\ Q_I 
  \end{matrix}
\right) 
- 2i \left(  
\begin{matrix} 
-\partial_{Q_I} \\ \partial_{P^I} 
  \end{matrix} 
\right) 
I_4^{1/2} (P^I, Q_I)~.
\end{equation}
Since the symplectic parametrization of the scalars fields is projective, the resulting physical coordinates on the scalar manifold are ratios 
$z^i = X^i/X^0$ for $i=1,\ldots, n_V$. 

The explicit determination of the function of charges $I_4(P^I, Q_I)$ requires solution of the attractor equations \eqref{eqn:attractpq} 
or, equivalently, extremization of the spacetime central charge over all moduli. This problem can be challenging for specific theories but it 
has been much studied so we will consider $I_4(P^I, Q_I)$ a known function of the charges. The goal in this work is to leverage the study of 
the extremal AdS$_2 \times S^2$ BPS attractors, including the computation of $I_4$, to illuminate also the nAttractor behavior of nearly extreme black holes. 

\subsection{Near Horizon Perturbation Theory}
\label{sec:nearhorizonIR}
The symmetry breaking scale $L$ of the near horizon geometry can be computed from the BPS solution according to \eqref{eqn:Lformula}. 
However, it depends on the derivative $\partial_r R$ at the horizon so it is not a property of the AdS$_2\times S^2$ attractor geometry
by itself. In other words, it depends on the full BPS flow equations \eqref{eqn:attractorflow} rather than just the attractor equations \eqref{eqn:attractpq}. 

The derivative $\partial_r R$ at the horizon is evidently a near horizon property so a natural strategy is to study the flow 
equations \eqref{eqn:attractorflow} in perturbation theory around the attractor geometry. 
However, given an AdS$_2\times S^2$ solution, expansion of \eqref{eqn:attractorflow} around $r=0$ 
does not determine derivatives like $\partial_r R$ and $\partial_r z^i$ at the horizon, even though the flow equations \eqref{eqn:attractorflow} are of first 
order. For example, the derivative $\partial_r R$ drops out at linear order and at quadratic order we find 
\begin{equation}
 \partial^2_r R   = - R g_{i{\bar\j}} \partial_r z^i \partial_r {\bar z}^{\bar\j}  ~, 
\label{eqn:d2R}
\end{equation}
which does not impose a useful constraint on the first derivatives. 
 
As a guide for expectations, we can treat the radial evolution of the fields $R, z^i$ as a mechanical system. In this analogy a well-posed initial value problem requires specification of both ``positions" $R, z^i$ and ``velocities" $\partial_r R$, $\partial_r z^i$ at 
the initial ``time". In fact, with the initial ``time" at the extremal horizon, the initial ``positions" $z^i$ are not actually arbitrary, they 
must be fixed at their attractor values. The mechanics problem is therefore somewhat degenerate, but the analogy with mechanics 
nonetheless demonstrates that the derivatives $\partial_r R$ and $\partial_r z^i$ at the horizon constitute additional input data that can be specified at will. 

In the context of nAdS$_2$/nCFT$_1$ holography each of the scalar derivatives $\partial_r R$, $\partial_r z^i$ specify dimensionfull 
parameters that must be represented in the dual nCFT$_1$. We can identify them with the coefficients of irrelevant operators that each 
break the scale invariance of AdS$_2$. This interpretation challenges the notion that ``the" dilaton corresponding to the radial function 
$R$ encodes a universal symmetry breaking scale. Generally the near horizon nCFT$_1$ can not be characterized by just one scale in any operational sense: there is no reason the additional data can not introduce a hierarchy of scales. 

\subsection{The BPS Flow}
\label{sec:nearhorizonUV}
The near horizon perturbation theory offers an IR perspective on the scalars in the nAdS$_2$ region: their slopes are parametrized 
by the derivatives of scalars flowing out from the attractor geometry. The corresponding UV description specifies the values of the 
scalars at infinity. Any value of these parameters specify uniquely, through the radial evolution of the BPS black hole, a 
nAttractor\footnote{We assume that charges and moduli have been specified so that the near horizon geometry is AdS$_2\times S^2$. Thus parameters are restricted so singular near horizon geometries are avoided, nor do we allow for wall crossing phenomena. The number of continuous parameters is independent of these limitations. (Recall that classical charges $(Q^I, P_I)$ are considered continuous parameters.)}. The explicit relation between asymptotic parameters and the nAdS$_2$ geometry depends on the complete solution to the BPS flow equations \eqref{eqn:attractorflow}. Fortunately, this BPS flow has been known for a long time \cite{Denef:2000nb,Bates:2003vx}. 

It is convenient to introduce the asymptotic values of the moduli $z^i$ in their symplectic incarnation $(X^I_\infty, F^\infty_I)$, as we did for scalars at the horizon. Recall that this parametrization of the scalar fields is projective so the physical coordinates on the scalar manifold are ratios. Alternatively, we can 
identify physical scalars by picking a particular gauge such as $X^0=1$. Either way, the $2n_V+2$ components $(X^I_\infty, F^\infty_I)$ 
with $I=0,\ldots,n_V$ are equivalent to the $n_V$ complex scalar fields $z^i$ with $i=1,\ldots, n_V$. In the following we will retain the 
projective form of the scalars and not impose a gauge condition. 

The map \eqref{eqn:PQtoXFmap} yields the fixed horizon values of the scalars $(X^I_{\rm hor}, F^{\rm hor}_I)$ in terms of asymptotic 
charges $(Q^I, P_I)$. We use the inverse map to characterize the asymptotic scalars $(X^I_\infty, F^\infty_I)$ in terms
of variables $(p^I_\infty , q^\infty_I )$
\begin{equation}
\left(  
\begin{matrix} 
  X^I_\infty \\ F_I^\infty
  \end{matrix}
\right) = 
\left(  
\begin{matrix} 
  p^I_\infty \\ q^\infty_I 
  \end{matrix}
\right) 
- 2i \left(  
\begin{matrix} 
-\partial_{q^\infty_I} \\ \partial_{p^I_\infty} 
  \end{matrix} 
\right) 
I_4^{1/2} (p^I_\infty, q_I^\infty)~.
\label{eqn:qipidef}
\end{equation}
The symplectic section $(p^I_\infty , q^\infty_I)$ is analogous to the charges but $(P^I , Q_I)$ but it is {\it dimensionless} and, since it parametrizes the asymptotic scalars of a black hole, it is not related to any conserved current. Since the asymptotic data $(X^I_\infty, F^\infty_I)$ is projective, the analogue ``charges" $(p^I_\infty , q^\infty_I)$ are not unique. They must be normalized so that $I_4(p^I_\infty , q^\infty_I )=1$ in order that the black hole introduced below is asymptotically flat with properly normalized metric. They must also satisfy the orthogonality constraint $p^I_\infty q_I - p^I q^\infty_I = 0$. Thus the $n_V$ complex parameters $z^i_\infty$ are encoded in the
$2n_V+2$ real parameters $(p^I_\infty , q^\infty_I)$ that are subject to two real constraints. 
 
It is remarkable that in this framework the equation \eqref{eqn:R4att} for the scale $R(0)$ of the AdS$_2\times S^2$ 
attractor geometry as a function of the charges $(P^I, Q_I)$ also gives the functional form of the radial function 
$R$ that solves the full flow equations \eqref{eqn:attractorflow}. It is \cite{Bates:2003vx}: 
\begin{equation}
R^4(r) = I_4(P^I + p^I_\infty r, Q_I + q^\infty_I r)~.
\label{eqn:R4flow}
\end{equation}
The specific heat follows as a modest corrollary of this solution. Equivalently, we find the symmetry breaking 
scale \eqref{eqn:Lformula}:
\begin{equation}
L =   {2\pi\over G_4} \left( \partial_r R^4\right)_{\rm hor}
= {2\pi\over G_4}  \left( p^I_\infty {\partial\over\partial P^I} + q_I^\infty {\partial\over\partial Q_I} \right) I_4(P^I, Q_I)~.
\label{eqn:Leformula}
\end{equation}
This expression amounts to simple practical prescription. To reiterate: the
dependence of $I_4$ follows from entropy extremization (or other methods) applied to the AdS$_2\times S^2$ attractor geometry of the BPS 
black hole; and then \eqref{eqn:Leformula} gives the symmetry breaking scale of the near horizon theory by simple differentiation. The 
derivatives are with respect to charges which in the present context are continuous parameters. Indeed, they are the only parameters 
that the attractor geometry depend on. 

We can analyze the scalar moduli $z^i$ similarly. Their fixed values at the horizon were given as functions of the 
charges $(P^I, Q_I)$ in \eqref{eqn:PQtoXFmap}. The important point is that the radial dependence of the
scalars that supports the complete BPS flow solution has the same functional dependence on charges. Thus
it is given by the same substistituion that gave the radial function \eqref{eqn:R4flow}, wiz.
\begin{equation}
z^i(r) = z^i_{\rm hor}(P^I + r p^I_\infty, Q_I+ rq^\infty_I)~.
\label{eqn:zihor}
\end{equation}
Therefore, we can again present the radial derivative at the horizon in terms of derivatives that act on the space of charges 
\begin{equation}
\left({dz^i\over dr} \right)_{\rm hor}
= \left( p^I_\infty {\partial\over\partial P^I} + q_I^\infty {\partial\over\partial Q_I} \right) z^i_{\rm hor}(P^I, Q_I) ~.
\label{eqn:ziformula}
\end{equation}
The horizon values of the scalars $z^i_{\rm hor}$ are given by \eqref{eqn:PQtoXFmap} and we appealed to these expressions 
implicitly in the argument above. However, practical computations it may well be simpler to obtain these values by an extremum 
condition such $\partial_i |{\cal Z}|=0$ or by solving the attractor equations \eqref{eqn:attractpq}. The substitution 
rule \eqref{eqn:zihor} and the formula \eqref{eqn:ziformula} for the radial derivative apply without regard to the provenance of
the functional dependence of the horizon scalars $z^i_{\rm hor}$ on the asymptotic charges $(P^I, Q_I)$.

\subsection{Discussion}
The nAdS$_2/CFT_1$ approach to AdS$_2$ quantum gravity relies on effective quantum field theory so, as discussed in the introduction, the scales that enter play a central role. The formulae derived in the previous subsection offer some perspectives. 

For a generic extremal black hole with AdS$_2$ scale $\ell_2$, we expect all charges $(P^I, Q_I)\sim \ell_2$ and, at a generic point in moduli space, all the scalars $p^I_\infty, q_I^\infty\sim 1$. With these estimates the formula \eqref{eqn:Leformula} for the symmetry breaking scale $L$ gives $L\sim L_{\rm typ}$ where  $L_{\rm typ} = 8\pi \ell^3_2/G_4$ was introduced in \eqref{eqn:genericL}. However, a specific black hole solution will depend on the precise charges and moduli so even though the scale $L$ is generically of order $L_{\rm typ}$, it will not generally take the precise value \eqref{eqn:genericL}, with these factors of $2$ and $\pi$. More importantly, it is possible to tune the values of charges and/or moduli away from their generic values so that the symmetry breaking scale differs parametrically from $L_{\rm typ}$. 
 
Now, this discussion is based on identification of the inverse massgap as ``the" symmetry breaking scale $L$ while generally many more scales are introduced by the variation of all the scalar fields. In our conventions the radial function $R$ has dimension of length and the moduli are dimensionless so for ease of comparison we introduce the dilaton $\phi=R^2/\ell_2^2$ which has radial derivative at the horizon 
\begin{equation}
\left({d\phi\over dr} \right)_{\rm hor}
= \left( p^I_\infty {\partial\over\partial P^I} + q_I^\infty {\partial\over\partial Q_I} \right) \phi_{\rm hor}(P^I, Q_I) ~. 
\label{eqn:Phir}
\end{equation}
This formula is entirely analogous to \eqref{eqn:ziformula} for the scalar fields $z^i$. Therefore, the distance away from the AdS$_2\times S^2$ attractor needed for a relative deformation of a scalar by order unity is comparable for all these fields, and generically of order $\ell_2$. Of course the deformations must be small, they cannot be unity, so the near horizon region is much smaller than $\ell_2$. The important point here is that generically all the scalars deform the near horizon geometry comparably, although their detailed dependence on charges and moduli differ. 

The symmetry breaking scales characterizing the dual nCFT$_1$ are vacuum expectation values (VEVs) of operators that are dual to the bulk scalars. The holographic map computes VEVs by variation of the on-shell action with respect to bulk sources. Since all scalars generically have comparable impact on the bulk geometry they correspond to similar sources and so their dual VEVs will also be comparable. These VEVs will generally be large when the gravitational coupling is weak because they are proportional to the on-shell action which is estimated by the ground state entropy $S_0 = \pi\ell_2^2/G_4$. Inspired by the
inverse massgap \eqref{eqn:Leformula} we {\it define} precise length scales in the nCFT$_1$ as the inverse lengths in \eqref{eqn:Phir} and \eqref{eqn:ziformula}, multiplied by a large universal factor:
\begin{eqnarray}
L & =&  {4\pi\ell^4_2\over G_4} \left({d\ln \phi\over dr} \right)_{\rm hor}~,\cr
L^i & = & {4\pi\ell^4_2\over G_4} \left({d\ln z^i\over dr} \right)_{\rm hor}~.
\label{eqn:LLI}
\end{eqnarray}
We expect that computation of the VEVs through holographic renormalization would give similar results. 

The scales $L^i$ are defined in \eqref{eqn:LLI} as complex numbers. In this work we have made no effort to preserve supersymmetry through the spherical reduction. In a fully supersymmetric theory, the scale $L$ would similarly acquire a pseudoscalar partner and become complex. The ${\cal N}=2$ version of the SYK model indeed features a partner $\sigma$ to the scale $R^2$ \cite{Fu:2016vas}. The most precise way of counting scales may well be to refer to $n_V+1$ real scales and $n_V+1$ phases. 

The minimally coupled fields that are often invoked as probes in studies of nAdS$_2/CFT_1$ can be interpreted in the ${\cal N}=2$ SUGRA formalism as components of hypermultiplets. They are true moduli so they are massless which, according to \eqref{eqn:hform}, corresponds to operators with conformal weight $h=1$. In contrast, the quadratic fluctuations of the radial function $R^2$ and the complex scalar fields $z^i$ in the AdS$_2$ attractor geometry all satisfy a Klein-Gordon equation with $m^2 = 2\ell_2^{-2}$. Therefore, according to \eqref{eqn:hform} they all correspond to operators in the dual theory that have conformal weight $h=2$. Diffeomorphism invariance imposes additional constraints on $R^2$ that removes it as a propagating field in the 2D bulk, leaving only a 1D boundary field. Despite this difference, all these fields are put an a similar footing by their common conformal weight.

\section{Example: the STU Model}
In this section we apply the formulae for symmetry breaking scales to the ``four-charge" black holes in the STU model. This clarifies concepts introduced abstractly in the previous section \ref{sec:bhsols}. The relation of the STU model to Jackiw-Teitelboim gravity was recently studied in \cite{Li:2018omr}.

\subsection{The STU Model}
The STU model has $n_V=3$ ${\cal N}=2$ vector multiplets and the prepotential 
\begin{equation}
F = {X^1 X^2 X^3 \over X^0}~,
\end{equation}
where $X^I$ with $I=0,1,2,3$ are the four projective scalars. They correspond to three complex physical scalars $z^i = X^i/X^0 = x^i - i y^i$ with $i=1,2,3$. 

The four gauge fields in the model allow $8$ charge parameters $(p^I, q_I)$ with $I=0,1,2,3$. The general black hole solution with $8$ arbitrary charges and $3$ complex moduli is known \cite{Chow:2013tia}. However, for the purposes of a pedagogical example it is preferable to specialize to the ``four-charge" model where $p^0=q_i=0$. This simplified model has $4$ non-trivial charge parameters $q_0, p^i$ with $i=1,2,3$. We will find that with the restricted charge assignment and our phase choices, $X^i$ (and $F_0$) are purely real and $X_0$ (and $F^i$) are purely imaginary. The physical scalars $z^i = X^i/X^0 = x^i - i y^i$ therefore become purely imaginary. 

We can interpret the STU-model as the low energy limit of IIA string string on the six-torus $T^6 = T^2 \times T^2 \times T^2$. 
We pick the duality frame so $q_0$ is the number of $D0$ branes and 
the $p^i$ are the number of $D4$-branes wrapped on the $T^2\times T^2$ that is complementary to the $i$th $T^2$. The $p^0$ would correspond 
to $D6$-brane number and the $q_i$ to $D2$-branes on $T^2$ but the simplified charge configuration has taken 
these to vanish. The supersymmetry condition are consistent with signs so $q_0>0, p^i>0$. The three real moduli $y^i$ (with $i=1,2,3$) that are active with the simplified charge assignments can be identified with the volumes of the three tori in string units: $y^i= V^{(i)}_2/(2\pi\sqrt{\alpha'})^2$. 

The spacetime central charge \eqref{eqn:Zdef} with K\"{a}hler potential defined in \eqref{eqn:kahlerdef} becomes 
\begin{equation}
{\cal Z} = {1\over\sqrt{G_4}} e^{{\cal K}/2}\left( X^0 q_0 - F_i p^i\right) = {i\over\sqrt{8G_4 y^1 y^2 y^3}} ( q_0 + p^1 y^2 y^3 + p^2 y^3 y^1 + p^3 y^1 y^2 ) ~.
\label{eqn:styZ}
\end{equation}
for the four-charge configurations in the STU-model. The BPS mass is given by the norm of the spacetime central charge $|{\cal Z}|$ \eqref{eqn:styZ} evaluated at its asymptotic value. To confirm this, recall that the gravitational coupling is related to the six-volume in string units $v_6 = V_6/(2\pi\sqrt{\alpha'})^6 
= y_\infty^1 y_\infty^2 y^3_\infty$ {\it at infinity} through
\begin{equation}
{1\over\sqrt{8 G_4 v_6}} = {1\over g_s \sqrt{\alpha'}}~, 
\label{eqn:G4v}
\end{equation}
and the mass of a single $D0$ brane is $1/R_{11} = 1/(g_s \sqrt{\alpha'})$. The asymptotic value of the first term in \eqref{eqn:Zdef} thus gives the 
mass of $q_0$ $D0$-branes and the three remaining terms give analogous contributions from the $p^i$ D4-branes that wrap the $T^2\times T^2$ complementary to the $i$th
$T^2$.

\subsection{Computation of the Symmetry Breaking Scale}
The simplest incarnation of the attractor mechanism for an extremal black hole is the condition \eqref{eqn:DiZ} that the fixed values of moduli at the horizon extremize the norm of the spacetime central charge. Extremization of \eqref{eqn:styZ} over $y^i$ with $i=1,2,3$ gives
\begin{equation}
y^i_{\rm hor} = \sqrt{ q_0   \over p^1 p^2 p^3}~p^i~. 
\label{eqn:y1hor}
\end{equation}
The central charge evaluated at this extremum then gives
\begin{equation}
I_4(P^I,Q_I) = R^4 = G^4_4 |{\cal Z}_{\rm hor}|^4  = 4 G^2_4 q_0 p^1 p^2 p^3 ~,
\label{eqn:I4stu}
\end{equation}
corresponding to the standard expression for the black hole entropy of the four-charge BPS black hole
\begin{equation}
S = {\pi\over G_4} R^2 =  2\pi\sqrt{q_0 p^1 p^2 p^3} ~.
\end{equation}
As a consistency check, note that we can recover the horizon values for the scalars \eqref{eqn:y1hor} by inserting the result for $I_4$ \eqref{eqn:I4stu} in the general formula \eqref{eqn:PQtoXFmap}.

Before computing the symmetry breaking scales of the theory we also need to specify the asymptotic value of the scalars. 
In our parametrization the moduli are given as $y^i_\infty$ with $i=1,2,3$ but, according to \eqref{eqn:qipidef}, we must represent them as 
analogue ``charges" $q_0^\infty, p^i_\infty$ that are such that the moduli satisfy the asymptotic analogues of their attractor values \eqref{eqn:y1hor}, wiz. 
\begin{equation}
y^i_\infty = \sqrt{ q_0^\infty   \over p^1_\infty p^2_\infty p^3_\infty}~p^i_\infty~. 
\label{eqn:y1infty}
\end{equation}
Solution of these equations give the ``charges" $q_0^\infty, p^i_\infty$ up to a common factor which is determined 
by the normalization condition $I_4 (p^I_\infty,q_I^\infty) = q^\infty_0 p^1_\infty p^2_\infty p^3_\infty = 1$. This inversion of \eqref{eqn:y1infty} gives
\begin{equation}
q^\infty_0 = \sqrt{y_\infty^1 y_\infty^2 y_\infty^3}~,
\end{equation}
and
\begin{equation}
p^i_\infty = {1\over\sqrt{y_\infty^1 y_\infty^2 y_\infty^3}}~y^i_\infty~. 
\end{equation}
Comparison with the spacetime central charge \eqref{eqn:styZ} shows that the dimensionless parameters $q_0^\infty, p^i_\infty$ have the
same dependence on moduli as the {\it inverse} mass of the corresponding type of brane.
 
After these preparations, it is now straightforward to compute the symmetry breaking scale \eqref{eqn:Leformula} 
\begin{eqnarray} 
L  &=& {2\pi\over G_4} \left( p^i_\infty {\partial\over\partial P^i} + q_0^\infty {\partial\over\partial Q_0} \right) R^4 \cr
&=& 2\pi  q_0 p^1 p^2 p^3 \sqrt{8G_4} \left( {q_0^\infty\over q_0}  +{p^1_\infty\over p^1}  +{p^2_\infty\over p^2} +{p^3_\infty\over p^3} \right) \cr
&=& 2\pi q_0 p^1 p^2 p^3 R_{11} ( {1\over q_0} + {1\over p^1 y_\infty^2 y_\infty^3}+ {1\over p^2 y_\infty^3 y_\infty^1} + {1\over p^3 y_\infty^1 y_\infty^2}  )~.
\label{eqn:Lresult}
\end{eqnarray}
In the evaluation we related dimensionless (quantized) charges $q_0, p^i$ and ``physical" charges $Q_0, P^i$ with dimension length
by the universal factor $\sqrt{2G_4}$. We used \eqref{eqn:G4v} for Newton's constant $G_4$ to introduce the radius of the M-theory circle $R_{11} = g_s \sqrt{\alpha'}$ in the final line. The parametrization of asymptotic scalars by moduli ``charges" $q_0^\infty, p^i_\infty$ appeared only at the intermediate stage of the computation and was ultimately eliminated in favor of the moduli $R_{11}$ and $y^i_\infty$ that have geometrical significance in higher dimensions. 

\subsection{Discussion}
We presented the final answer \eqref{eqn:Lresult} for the symmetry breaking scale of the four charge black hole in a form where it is reminiscent of the long string scale for this model 
\begin{equation}
L_{\rm long} = 2\pi p_1 p_2 p_3 R_{11}~. 
\label{eqn:longstr}
\end{equation}
This is interesting because this scale (or at least minor variations of this scale) also appears as the inverse massgap in important microscopic models of black holes, such as the MSW model \cite{Maldacena:1997de}.

To make this connection more precise, recall that the contribution to the total black hole mass from the $q_0$ D0-branes is $q_0/R_{11}$. The remaining three terms in the symmetry breaking scale \eqref{eqn:Lresult} are similarly related to the mass of the D4-branes (of type $i$) such that each of the four term is proportional to the {\it inverse} of a mass contribution. The scale $L$ reduces precisely to the long string scale $L_{\rm long}$ when moduli are tuned so that the $D0$-branes are much lighter than the D4-branes. This is the ``dilute gas" limit where the D0-branes can be treated as excitations of the D4-branes and it is in this regime that the microscopic description 
yielding $L_{\rm long}$ is justified \cite{Maldacena:1996ix}. At a generic point in moduli space the four contributions to the symmetry breaking scale \eqref{eqn:Lresult} are comparable but it is reasonable to interpret this more general scale $L$ as an effective long string scale anyway. Indeed, this more symmetric generalization of the long string scale was extracted from black hole greybody factors already in \cite{Cvetic:1997xv} (where it was denoted ${\cal L}$). 

An alternative and model-independent characterization of the dilute gas limit is that it isolates the situations where the AdS$_2$ spacetime can be lifted consistently to an AdS$_3$ geometry. These special cases are important because in precisely this case the nCFT$_1$ can be constructed from a sector of the CFT$_2$ that is dual to the AdS$_3$ theory \cite{Gupta:2008ki,Balasubramanian:2009bg,Castro:2010vi}.

The Reissner-Nordstr\"{o}m black hole solution to Einstein-Maxwell theory (with a single vector field and no scalars) is the special case of the STU black hole where the mass contribution from the four charges are identical. In this case, the symmetry breaking scale \eqref{eqn:Lresult} reduces to the ``typical scale" \eqref{eqn:genericL} which is expressed entirely in terms of the horizon radius $\ell_S$, Newton's coupling constant, and numerical factors (chosen so it agrees with $L$ for the Reissner-Nordstr\"{o}m black hole). 

However, generically the symmetry breaking scale \eqref{eqn:Lresult} depends on the moduli of the theory. The dependence introduced in each of the four terms in the symmetry breaking scale \eqref{eqn:Lresult} is genuinely independent. Indeed, the scales introduced by the variation of the other scalar fields in the near horizon region \eqref{eqn:ziformula} gives
\begin{eqnarray} 
L^1 &=&  2\pi q_0 p^1 p^2 p^3 R_{11}  \left( {1\over q_0}   + {1\over p^1 y^\infty_2 y^\infty_3 } -
{1\over p^2y^\infty_1 y^\infty_3}-  {1\over p^3 y^\infty_1 y^\infty_2}\right)~, 
\label{eqn:LiSTU}
\end{eqnarray}
and formulae with $(123)$ permuted cyclically. It is clear that, by taking linear combinations, we can treat each of the four terms in the formulae as independent scales. 

The dilute gas black holes are special because for those the first term in \eqref{eqn:LiSTU} dominates and so the three scales $L^i$ are identical to ``the" symmetry breaking scale \eqref{eqn:Lresult}. Therefore, in this case the scales $L^i$ do not play an independent role. On the other hand, in this limit the symmetry breaking scale is much larger than the typical scale \eqref{eqn:genericL}. The Reissner-Nordstr\"{o}m black holes are special in a different way. For  those the four terms in the parenthesis have equal magnitude, and the signs are such that they cancel. Thus the ``new" scales $L^i$ vanish. In the dual nCFT$_1$ this corresponds to a sector of the theory becoming heavy, so that it decouples. However, generically all these scales must be realized in the dual CFT$_1$.

\section*{Acknowledgements}
We thank Alejandra Castro, Anthony Charles, Junho Hong, Jim Liu, Ioannis Papadimitriou, and Yangwenxiao Zeng for useful discussions and concurrent collaboration on related research. Preliminary versions of this work was presented at the Great Lakes Strings Conference 2018 in Chicago (April 2018) and at ``Rencontres Th\'{e}oriciennes" in Paris (May 2018). We thank participants for questions and comments. 
This work was supported in part by the U.S. Department of Energy under grant DE-FG02-95ER40899.

\appendix
\section{Special Geometry}
\label{app:spec}

{\it Special geometry} controls several aspects of ${\cal N}=2$ supergravity, such as the geometry of the manifold parametrized by the scalar 
fields and the gauge kinetic function. This appendix collects a few basic notions of special geometry. Some good references
for this material are \cite{deWit:2002vz,Freedman:2012zz}.

A special K\"{a}hler manifold allows a {\it symplectic section} $(L^I, M_I)$ with $I=0,\ldots, n_V$ that transforms in the 
fundamental representation of the duality group $Sp(2n_V+2)$. The {\it embedding} K\"{a}hler manifolds 
needed in supergravity possess a scaling symmetry so that, without loss of generality, we can impose the 
projection constraint
$$ 
i ({\bar L}^I M_I - L^I {\bar M}_I)=1~.
$$
This constraint is solved by the {\it holomorphic section} $(X^I, F_I)$ introduced through 
\begin{eqnarray}
L^I &=& e^{{\cal K}/2}X^I~, \cr
M_I &=& e^{{\cal K}/2}F_I~,
\label{eqn:xifihol}
\end{eqnarray}
where the K\"{a}hler potential ${\cal K}$ is defined by
\begin{equation}
{\cal K} = - \ln i ({\bar X}^I F_I - X^I {\bar F}_I)~.
\label{eqn:kahlerdef}
\end{equation}
As the terminology indicates, the holomorphic section $(X^I, F_I)$ is a {\it holomorphic} function of the coordinates $z^i$ on the projected K\"{a}hler 
space:
\begin{equation}
\partial_{\bar\i}X^I  = \partial_{\bar\i}F_I=0~.
\end{equation} 
The symplectic section $(L^I, M_I)$ is holomorphic on the embedding K\"{a}hler manifold but, because the projection \eqref{eqn:xifihol} involves 
the real K\"{a}hler potential ${\cal K}$, it is not holomorphic on the projected manifold. It is the latter that is the physical target space for the 
$\sigma$-model of scalars.

Derivatives of the K\"{a}hler potential ${\cal K}$ \eqref{eqn:kahlerdef} define a connection on the projective space. The spacetime central charge 
\begin{equation}
{\cal Z} = {1\over\sqrt{G_4}} e^{{\cal K}/2}\left( X^I q_I - F_I p^I\right)~,
\label{eqn:Zadef}
\end{equation}
is holomorphically {\it covariant}:  
\begin{eqnarray}
D_{\bar\i} {\cal Z}= (\partial_{\bar\i}- {1\over 2} \partial_{\bar\i}{\cal K}){\cal Z} & = & 0~,
\end{eqnarray}
with respect to this connection. The dependence of the spacetime central charge on the position in 
moduli space is therefore encoded in the holomorphic covariant derivative
\begin{equation}
D_i {\cal Z}  = (\partial_i +{1\over 2} \partial_i{\cal K}){\cal Z}  = 2 \sqrt{{\cal Z}\over{\overline{\cal Z}}}\partial_i  |{\cal Z}|~.
\end{equation}
The resulting identity 
\begin{equation}
g^{i{\bar\j}}D_i {\cal Z} \overline{D}_{\bar\j} \overline{\cal Z}  = 4g^{i{\bar\j}}\partial_i |{\cal Z}| \overline{\partial}_{\bar\j} |{\cal Z}| ~, 
\end{equation}
is useful for manipulations of the spacetime potential \eqref{eqn:Vpot}.

The projected manifold inherits the K\"{a}hler metric 
\begin{equation}
g_{i{\bar\j}} = \partial_i\partial_{\bar\j}{\cal K}~,
\label{eqn:gijK}
\end{equation} 
from the embedding manifold. The expression is invariant under the K\"{a}hler transformations 
$K(z^i,{\bar z}^i)\to K(z^i,{\bar z}^i) + f(z) + {\bar f} ({\bar z})$ for any complex $f$ (and its conjugate ${\bar f}$). 
The $U(1)$ line bundle over the projected manifold defined by this symmetry transformations has a non-trivial field strength 
that is also given by expression \eqref{eqn:gijK}. This relation is the hall-mark of a K\"{a}hler-Hodge manifold. 

The base coordinates $z^i$ and the fibre coordinate together form adequate coordinates on the embedding space. The 
diffeomorphism to the defining coordinates $X^I$ on the embedding manifold shows that the $(n_V+1)\times (n_V+1)$ matrix 
$(\overline{X}_J ~\nabla_i X_J)$ is invertible. This matrix enters the gauge kinetic terms
\begin{equation}
\mu_{IJ} - i \nu_{IJ}  = \overline{\cal N}_{IJ} = ( \overline{F}_I ~\nabla_i F_I)(\overline{X}_J ~\nabla_i X_J)^{-1}~.
\label{eqn:barnijsym}
\end{equation} 
The fully holomorphic $(n_V+1)\times (n_V+1)$ matrix $(X_J ~\nabla_i X_J)$ may similarly be invertible but it does not have to be. 
Invertibility of this matrix is the integrability condition for the existance of a prepotential $F(X^I)$ that is
homogeneous of degree two in its variables and generates the lower components of the sympletic vector $(X^I, F_I)$ through 
$F_I = \partial_I F$. When a prepotential exists the gauge kinetic terms \eqref{eqn:barnijsym} can be recast as 
\begin{equation}
\mu_{IJ} + i \nu_{IJ} = {\cal N}_{IJ} = \overline{F}_{IJ} + 2i { ({\rm Im} F_{IK}) X^K ({\rm Im} F_{JL}) X^L\over  X^M ({\rm Im} F_{MN}) X^N}~.
\end{equation} 
where $F_{IJ} = \partial_i \partial_J F$. 

\section{Equations of Motion}
\label{app:eom}
In this Appendix we analyze the equations of motion for the 2D theory with Lagrangian \eqref{eqn:2dlag}.

At the outset we partially fix the gauge so that the 2D metric is diagonal
\begin{equation}
ds^2_2 = - e^{2\Phi} dt^2 + e^{2\Psi} dr^2 ~.
\label{eqn:genmet}
\end{equation}
After we find the equations of motion we will fix the gauge fully by imposing also $\Psi=-\Phi$. For a static solution the 2D curvature is
\begin{equation}
{\cal R}^{(2)} =  - e^{-\Psi - \Phi} \partial_r \left( e^{-\Psi-\Phi} \partial_r e^{2\Phi} \right)~, 
\end{equation} 
and the 2D action \eqref{eqn:2dlag} simplifies to
\begin{eqnarray}
4G_4 {\cal L}_2 &=& e^{\Phi+\Psi} \left[ - R^2 e^{-\Psi - \Phi} \partial_r \left( e^{-\Psi-\Phi} \partial_r e^{2\Phi} \right)  + 2 
+ 2e^{-2\Psi} (\partial_r R)^2 - 2 R^2 e^{-2\Psi} g_{i{\bar\j}} \partial_r z^i \partial_r {\bar z}^{\bar\j}  - {2V_{\rm eff}\over R^2}  \right]~.\nonumber
\end{eqnarray}
Variation of the independent fields $R, \Phi, \Psi, {\bar z}_{\bar\j}$ then gives the equations of motion
\begin{eqnarray}
0 & =& - R^2 \partial^2_r e^{2\Phi}  -  2 R \partial_r \left( e^{2\Phi} \partial_r R\right)  - 2  R^2 e^{2\Phi}g_{i{\bar\j}} \partial_r z^i \partial_r {\bar z}^{\bar\j}  + {2V_{\rm eff}\over R^2}  ~,\cr
 0 &=&  - 2e^{\Phi}  
\partial_r  \left(  e^{\Phi} \partial_r R^2 \right) + 2 + 2e^{2\Phi} (\partial_r R)^2   - 2 e^{2\Phi} R^2
g_{i{\bar\j}} \partial_r z^i \partial_r {\bar z}^{\bar\j}   - {2V_{\rm eff}\over R^2} ~, \cr
0 &=&  - (\partial_r R^2) (\partial_r e^{2\Phi}) + 2 - 2e^{2\Phi} (\partial_r R)^2 + 2  e^{2\Phi} R^2 g_{i{\bar\j}} \partial_r z^i \partial_r {\bar z}^{\bar\j}  - {2V_{\rm eff}\over R^2} ~,
 \cr
0 &=& \partial_r \left(  R^2 e^{2\Phi} g_{i{\bar\j}}  \partial_r z^i  \right) - {1\over R^2}  \partial_{\bar\j} V_{\rm eff} ~, 
\label{eqn:alleom}
\end{eqnarray}
where we took $\Psi=-\Phi$. Note that the second and third of these equations are due to the independent variations of $\Phi$ and $\Psi$. If we had imposed 
the gauge condition $\Psi=-\Phi$ prematurely we would find only one linear combination of these two equations. 

The sum of the first and third equation in \eqref{eqn:alleom} involves neither the potential, nor the scalars. It can be rewritten as 
\begin{equation}
\partial^2_r (R^2 e^{2\Phi}) = 2~.
\end{equation}
Therefore we have $R^2 e^{2\Phi} = r^2$ up to a term that is linear in $r$. The slope of this linear term can be taken to vanish without loss of generality, by shifting the origin of $r$, and so the solution becomes
\begin{equation}
R^2 e^{2\Phi} = r^2-r^2_0~,
\label{eqn:RPHI}
\end{equation}
where $r^2_0$ is a constant of integration. 
This result shows that the 4D metric takes the form 
\begin{equation}
ds^2_4 = - {r^2-r^2_0\over R^2}  dt^2 +   {R^2\over r^2-r^2_0} dr^2 + R^2 d\Omega^2_2~,
\end{equation}
for any static solution. This restricted form is central for the analysis of nonextremal black holes. It plays a central role in section \ref{sec:nextbhs}.

The result \eqref{eqn:RPHI} simplifies the equations of motion \eqref{eqn:alleom}. We can recast them in the relatively simple form
\begin{eqnarray}
0 &=&  - \partial_r  \left(  {r^2-r^2_0\over R^2}\partial_r R^2 \right)  + 2  - {2V_{\rm eff}\over R^2}~, \cr
0 &=& {1\over R} \partial^2_r R  + g_{i{\bar\j}} \partial_r z^i \partial_r {\bar z}^{\bar\j}~,     \cr
0 &=& \partial_r \left(   (r^2-r^2_0) g_{i{\bar\j}}  \partial_r z^i  \right) - {1\over R^2}  \partial_{\bar\j} V_{\rm eff} ~. 
\label{eqn:simpleom}
\end{eqnarray}
The upper two equations are found by simplifying the sum and difference of the third and second equations in \eqref{eqn:alleom}. The last 
equation is the fourth equation in \eqref{eqn:alleom}, simplified using \eqref{eqn:RPHI}. All solutions to the BPS conditions \eqref{eqn:attractorflow} satisfy these
equations of motion. 


\providecommand{\href}[2]{#2}\begingroup\raggedright\endgroup

\end{document}